\documentstyle[preprint,version2,aps]{revtex}

\begin{document}
\draft
\begin{title}
Comment on "Microscopic Theory of Periodic Conductances \\
in Narrow Channels"
\end{title}
%%%%%%%%%%%%%%%%%%%%%%%
\author{Norio Kawakami}
%%%%%%%%%%%%%%%%%%%%%%%%
\begin{instit}
Yukawa Institute for
Theoretical Physics, Kyoto University, Kyoto 606, Japan.
\end{instit}
%\receipt{}
%\moreauthors{}
%\begin{abstract}
%\end{abstract}
\vspace{1 cm}    % for preprint style only
\narrowtext
%%%%%%%%%%%%%%%%%%%%%%%%%%%%%%%%%%%%%%%%%%%%%%%%%%%%%%%%%%%%%%%%%%%%
%%%%%%%%%%%%

In a recent letter Johnson and Payne (JP) have studied the
effect of the electron interaction on the
periodic conductance oscillations in narrow channels
based on an exactly solvable model, and
have explained several characteristic properties
observed experimentally  \cite{jp}.
They have obtained the successive peaks
in the oscillations with the single period for
{\it any} value of the interaction strength.
In their model, however, the exchange effect is completely neglected,
leading to the results essentially same as those of spinless fermions.
In this Comment we discuss the exchange effect on
JP's model using a solvable electron model.
We wish to point out that when the interaction strength
becomes rather small, there may appear two periods
in the conductance oscillations due to the exchange effect.

The model proposed by JP is the one-dimensional
continuum electron system confined by the harmonic potential
${1 \over 2} m^* \omega_0  \sum_i x_i^2$ with
repulsive interactions of the inverse-square type,
$\sum_{i>j} \lambda(\lambda+1)(x_i-x_j)^{-2}$, where the dimensionless
parameter $\lambda$ is expressed as
$\lambda=\tau-1/2$ with the parameter $\tau$ in JP's paper \cite{jp}.
Although this model is solvable it
completely neglects the exchange effect, so that the ground
state is highly degenerate due to the internal spin degrees
of freedom as is the case for the $U=\infty$
Hubbard model. A  solvable electron model
including the exchange effect has been proposed
by Polychronakos \cite{poly} and Ha-Haldane \cite{ha}
in a slightly different context.  The interaction takes the form
of $ \sum_{i>j} \lambda(\lambda+ P_{ij}^{\sigma}) (x_i-x_j)^{-2}$
where $P_{ij}^{\sigma}$ is the spin exchange operator of two
electrons.  This model enables us to
study how the exchange effect modifies the period of
the conductance oscillations in JP's model.
The ground-state wavefunction
includes the Vandermonde determinantal products \cite{jp},
the power of which is raised up to $\lambda+1$
for the particles with the same spins
and to $\lambda$ for the different spins \cite{ha}.
 Applying the analysis
of ref. \cite{ha} it can be deduced  that the energy increment
quadratically proportional to the electron number is written
in a succinct form,
%%%%%%%%%%%%%%%%%%%%%%%%%%%%%%%%
\begin{equation}
\Delta E (\Delta N)= {1 \over 2} \hbar \omega_0
\vec n^t {\bf A} \vec n,
\end{equation}
%%%%%%%%%%%%%%%%%%%%%%%%%%%%%%%
with the $2\times 2$ matrix $A_{ij}=\lambda +\delta_{ij}$,
where the integer-valued vector
$ \vec n=(n_\uparrow, n_\downarrow)$ labels
the change of the electron number as $\Delta N=n_\uparrow +
n_\downarrow $. The lowest-energy state associated with the
change of the electron number $(\Delta N=1, 2, 3, \cdots)$ is given
respectively by the quantum numbers
$\vec n= (1,0)$, $(1,1)$,  $(2,1)$, etc.
According to JP \cite{jp}  the spacing  $\delta$ of the
successive peaks in the conductance oscillations
is given by $\delta=\Delta E(N+2)+\Delta E(N)-2\Delta E(N+1)$,
which is evaluated as,
%%%%%%%%%%%%%%%%%%
\begin{equation}
\delta_1= \hbar \omega_0 \lambda,
\hskip 3mm
\delta_2= \hbar \omega_0 (\lambda +1).
\end{equation}
%%%%%%%%%%%%%%%%%
It is seen that there appear two periods for the conductance
oscillations reflecting  the internal spin degrees of freedom
in contrast to JP's model which is essentially
as same as  the spinless-fermion model (the spacing
is simply given by $\delta=\hbar \omega_0(\lambda+1)$) \cite{jp}.
For the parameters used by JP ($\delta \sim 8.5 \hbar \omega_0$),
we get from (2)  $\delta_1 \sim 7.5 \hbar \omega_0$ and
$\delta_2 \sim 8.5 \hbar \omega_0$, implying the correction due
to the exchange effect is rather small for these parameters.
As the interaction becomes weaker (smaller $\lambda$), however,
the above exchange effect becomes conspicuous, making two
periods more distinct. When the
interaction strength takes the vanishing value
($\lambda \rightarrow 0$), the present model reproduces the
results of free electrons in the harmonic potential
($\delta_1\rightarrow 0, \hskip2mm \delta_2
\rightarrow \hbar \omega_0$), which are consistent with the
results in ref. \cite{ha}.  Though the present model
includes the exchange effect in a specific way, it clearly
demonstrates that for weakly correlated cases
one may  naturally expect two kinds of the
oscillation periods by taking into account the
exchange effect even in a model with singular
interactions of the  $1/x^2$  type.

%\newpage
%**********************************************************
%******************** R E F E R E N C E S *****************
%*********************************************************


\begin{references}
%%%%%%%%%%%%%%%%%
\bibitem[1]{jp}
N. F. Johnson and M. C. Payne,
 Phys. Rev. Lett. {\bf 70}, 1513 (1993).
%%%%%%%%%%%%%%%%%%
\bibitem[2]{poly}
A. Polychronakos,
 Phys. Rev. Lett. {\bf 69}, 703 (1992).
%%%%%%%%%%%%%%%%%%%%%%%%%
\bibitem[3]{ha}
Z. N. C. Ha and F. D. M. Haldane,
Phys. Rev. B {\bf 46}, 9395 (1992).
%%%%%%%%%%%%%%%%%%%%%%%
\end{references}
\end{document}